\documentclass[pdflatex,sn-mathphys-num]{sn-jnl}

\usepackage{graphicx}%
\usepackage{multirow}%
\usepackage{amsmath,amssymb,amsfonts}%
\usepackage{amsthm}%
\usepackage{mathrsfs}%
\usepackage[title]{appendix}%
\usepackage{xcolor}%
\usepackage{textcomp}%
\usepackage{manyfoot}%
\usepackage{booktabs}%
\usepackage{algorithm}%
\usepackage{algorithmicx}%
\usepackage{algpseudocode}%
\usepackage{listings}%
\usepackage{float} 
\floatstyle{plain}
\newfloat{diagram}{htbp}{lop}[section]
\floatname{diagram}{Diagram}

\theoremstyle{thmstyleone}%
\newtheorem{theorem}{Theorem}
\newtheorem{proposition}{Proposition}
\newtheorem{assumption}[theorem]{Assumption}%

\theoremstyle{thmstyletwo}%
\newtheorem{example}{Proof}%

\theoremstyle{thmstylethree}%
\newtheorem{definition}{Definition}%
\usepackage{parskip}
\usepackage{changepage}   

\begin{document}

\title[Article Title]{Binary or nonbinary? An evolutionary learning approach to gender identity}


\author{\fnm{Hung} \sur{Truong} \email{hungt@sfu.ca}}



\affil{\orgdiv{Economics Department}, \orgname{Simon Fraser University}, \country{Canada}}




\abstract{Is gender identity binary or nonbinary? My analysis shows that while both are possible, the latter is a more attracting equilibrium under an adaptive learning perspective. I frame the gender identity problem as a modified \textit{battle of the sexes} game, where individuals define their gender identity under pairwise matching motives. From a baseline game-theoretical standpoint, I demonstrate that the binary-only world and the nonbinary-only world are both Nash equilibria in the stage game and are locally stable in the infinitely repeated game. Thus, any state of gender identity could theoretically persist. I then adopt a genetic learning algorithm as an equilibrium selection criterion to investigate evolutionary dynamics further and provide a rationale for the transition from binary to nonbinary gender identity. Specifically, in a binary-origin world, divergence occurs as individuals identifying as nonbinary gender evolve to become the majority due to their higher flexibility in matching outcomes. My framework captures how adaptive learning drives identity evolution, offering a parsimonious tool to analyze how diversity and exclusivity emerge in varying economic environments.
}




\maketitle
\section{Introduction}\label{sec:intro}

Understanding the evolution of social identities, including gender identity, is critical to explaining how diversity and exclusivity emerge in various economic and social contexts. While traditional economic models often treat identity as fixed or exogenous, a growing body of interdisciplinary research emphasizes that identity is malleable and influenced by social and economic interactions \citep{akerlof2000economics,shayo2020social}. In particular, gender identity—whether binary or nonbinary—raises fundamental questions about how individuals define themselves and interact within their environments. This paper contributes to this literature by providing a novel framework to analyze the dynamics of gender identity evolution and its implications for social coordination and diversity.

Social identity theory underscores the interplay between individual preferences and group affiliation, suggesting that identity formation is shaped by the incentives and trade-offs inherent in social and economic structures. In a recent survey of the literature, \cite{shayo2020social} discusses works on social identity and economic policy, and demonstrates how individuals’ identification with different social groups can drive preferences over redistributive policies. Similarly, the framework developed in this paper explores how individuals’ gender identity evolves in response to pairwise matching motives, reflecting both individual utility maximization and broader social dynamics.

The conceptual foundation of this paper lies in game theory, specifically a modified version of the classic \textit{battle of the sexes} game. The game theoretical setup captures the tension between coordination and individual preferences in defining gender identity. At a baseline game-theoretical standpoint, both binary and nonbinary gender identity worlds are Nash equilibria of the stage game, and are locally stable then the game is infinitely repeated. These findings align with the theoretical prediction that multiple equilibria can coexist, contingent on the social and economic environment \citep{young2020individual}.

To move beyond stage equilibrium analysis, this paper introduces a genetic learning algorithm \citep{holland1992genetic} as an equilibrium selection criterion, drawing inspiration from evolutionary game theory and adaptive learning models \citep{sandholm2010population}. The genetic algorithm models how individuals adapt and evolve their gender identities over time based on the payoffs from pairwise matching. The results reveal a striking transition dynamic: starting from a binary-origin world, the population evolves toward a predominance of nonbinary gender identity. This transition is driven by the higher flexibility and adaptability of nonbinary identities, which enhance matching outcomes and social coordination.

The broader implications of these findings are twofold. First, they provide a rationale for the observed increase in nonbinary gender identification in contemporary societies, highlighting the role of adaptive learning and social dynamics. Second, the framework offers a parsimonious tool for analyzing the emergence of diversity and exclusivity across different economic and social environments. By situating the analysis within the broader context of social identity theory and evolutionary dynamics, this study bridges the gap between theoretical modeling and real-world phenomena.

\paragraph{Related literature} 
This analysis builds on and contributes to a rich tradition of research on social identity, including foundational works by \cite{akerlof2000economics,chen2011potential,benjamin2010social,charness2020social,shayo2020social}. It also draws methodological inspiration from macroeconomic equilibria and adaptive learning models \citep{arifovic2000evolutionary, arifovic1997transition, arifovic2023history}, providing a novel lens to understand identity evolution in dynamic and diverse environments. Research on conformity to ingroup norms demonstrates that individuals adapt their behaviors to align with those of their ingroup, influenced by social salience, primed group membership, and shared threats \citep{abrams1990knowing,van2013self,lakin2003using,benjamin2010social,hoff2006discrimination}. This aligns with the framework in my study, highlighting how social identity dynamics shape adaptive learning and identity evolution. Notably, \cite{chen2011potential} present insights on how salient group identity fosters efficient coordination through learning, which corroborate my framework by illustrating the role of identity-driven dynamics in shifting equilibria, reinforcing the evolutionary transition from binary to nonbinary gender identity.

\medskip
\medskip

This paper is structured as follows. Section \ref{sec:framework} details the environment framework and the use genetic learning algorithm as what drive the learning process of individuals. Section \ref{sec:gametheory} introduces the theoretical model and outlines the game-theoretical properties of the stage game. Section \ref{sec:findings} presents the results from the genetic algorithm simulations and discusses the implications of the transition dynamics from binary to nonbinary gender identity. Finally, Section \ref{sec:conclusion} concludes by situating the findings within the broader literature on social identity and offering directions for future research.

\section{Framework}\label{sec:framework}
\subsection{The ``world"}
The computerized environment in this analysis consists of $N$ heterogeneous individuals. Each individual is represented by a chromosome encoded as a binary string of length $3 \times l$ genes/bits. For instance, with $l = 10$, a chromosome would be represented as:

\begin{equation*}
    \underbrace{010110111}_{\text {type }} \underbrace{0100010110}_\alpha \underbrace{1101010100}_\beta
\end{equation*}

\begin{figure}[h]
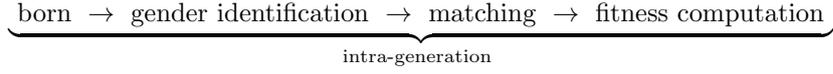

    \centering
    \begin{equation*}
    \underbrace{\text { born } \rightarrow \text { gender identification } \rightarrow \text { matching } \rightarrow \text { fitness computation }}_{\text {intra-generation }}
\end{equation*}
    \caption{Intra-generation activities}
    \label{fig:intragent}
\end{figure}

The first $l$ bits represent the “type” of an individual, which is inherently stochastic and subject to no evolutionary processes beyond randomness. In other words, the type of an individual is naturally determined by random initialization. The remaining $2 \times l$ genes form the biological interpreter, which translates the type into a corresponding gender identity. These $2 \times l$ genes undergo evolutionary processes that reflect natural selection, where gene strings associated with higher fitness are more likely to be selected for reproduction. I refer to these components as the type-chromosome, $\alpha$-chromosome, and $\beta$-chromosome strings. Each of these strings serves a distinct role in determining how an individual’s type is translated into gender identity. Figure \ref{fig:intragent} summarizes the processes that govern the interactives within the population.

First, individuals are born, and each identifies their gender. Matching then occurs, allowing individuals to obtain payoffs based on the matching outcomes. Specifically, as previously mentioned, an individual is born with a chromosome that consists of a stochastic type and a biological interpreter, which may be inherited from prior generations. For gender identification, the binary strings are decoded, normalized, and processed by the biological interpreter to produce a gender identity. Matches are then established according to specific rules, which I will outline in the following sections. The procedures for decoding gender identity are detailed below.

The strings are decoded as follows. First, the three chromosome parts are decoded to three base-two integers $m_{i,\{t y p e, \alpha, \beta\}}$, where the subscript $i$ indicates individual $i$,

\begin{equation}
    m_{i,\{t y p e, \alpha, \beta\}}=\sum_{k=1}^l a_{i,\{\text {type}, \alpha, \beta\}} 2^{k-1}
\end{equation}

Now, normalization. The   $type$-chromosome is normalized differently than   $\alpha$- and  $\beta$ -chromosome; and returns  $x_i\in [0,1]$, where  $x_i$  is the type of the individual. Specifically, the decoded and normalized   $type$-chromosome string takes the following expression

\begin{equation}
    x_i=\frac{m_{i, t y p e}}{\bar{K}_{ {type }}}
\end{equation}

$\bar{K}_{\text {type }}$ is the normalized parameter for $type$   and is selected such that $x_i\in [0,1]$. The normalized  $\alpha$- and   $\beta$-chromosome are

\begin{equation}
\alpha_i =\frac{m_{i, \alpha}}{\bar{K}_\alpha}
\end{equation}
\begin{equation}
    \beta_i =\frac{m_{i, \beta}}{\bar{K}_\beta}
\end{equation}

$\bar{K}_{\alpha,\beta}$ are the normalized parameters for  $\alpha$- and   $\beta$-chromosome. Let  $\xi_i$  be the gender identity of individual   $i$, I now illustrate the relationship between an individual’s type and its gender identity. I first define
\begin{equation}\label{eq:behav_int}
    x_i \equiv F\left(\xi_i, \alpha_i, \beta_i\right)
\end{equation}

$F(\cdot)$ is an invertible mapping heuristic. It follows that once $i$’s genetic trait pins down $x_i$, $\alpha_i$, and $\beta_i$, gender identity $\xi_i$ of individual $i$ can be obtained by 
\begin{equation}
    \xi_i=F\left(\xi_i, \alpha_i, \beta_i\right)^{-1}
\end{equation}

$F(\cdot)^{-1}$ is the biological interpreter. Given an individual’s type $x$ and a specific $F(\cdot)$, the individual translates its type into gender identity, to which the support of $\xi$ is the gender set. I now present the key specification of $F(\cdot)$:
\begin{equation}\label{eq:keyF}
    F\left(\xi_i, \alpha_i, \beta_i\right)=B\left(\frac{1}{\lambda\left(\alpha_i\right)}, \frac{1}{\lambda\left(\beta_i\right)}\right) \int_0^{\xi_i} t^{\frac{1}{\lambda\left(\alpha_i\right)^{-1}}}(1-t)^{\frac{1}{\lambda\left(\beta_i\right)}-1} d t=x_i
\end{equation}

$F(\cdot)$ takes the form of the cumulative distribution function of the \text{Beta} distribution with shape parameters $\frac{1}{\lambda\left(\alpha_i\right)}$ and $\frac{1}{\lambda\left(\beta\right)}$ (i.e., $\xi \sim \text{Beta}(\frac{1}{\lambda\left(\alpha_i\right)},\frac{1}{\lambda\left(\beta\right)})$ ), and $\frac{1}{\lambda\left(\alpha_i\right)}$ and $\frac{1}{\lambda\left(\beta\right)}$ describes the influence of $\alpha$ and $\beta$ on the shape of the distribution (details and properties are in appendix A). $F(\cdot)^{-1}$ is therefore the inverted cumulative distribution function that maps $x_i, \alpha_i$, and $\beta_i$ to the gender set, i.e., the support of $\xi$. I employ the \text{Beta} distribution due to its flexible data-generating process in the support $[0,1]$, making it well-suited for modeling the gender spectrum. While \text{Beta} distribution is generally a continuous probability distribution defined on $[0,1]$ (e.g., gender spectrum), it can be modified to generate binary outcomes; hence gender identity needs not be strictly binary nor nonbinary.  It is worthwhile to emphasize that an individual’s genetic trait, i.e., $\alpha$-chromosome and $\beta$-chromosome, has control over the specific form of its biological interpreter Equation \eqref{eq:behav_int}, and hence determines its potential gender set to be binary, $\xi \in \{0,1\}$; or nonbinary, $\xi \in (0,1)$. Conveniently, by imposing restrictions on the initial chromosome pools, a binary-origin environment can be established to explore whether divergence occurs.

\subsection{Matching}\label{sec:matching}
In this environment, matching is the primary objective of every individual, as it determines their payoffs. Matching occurs in pairs and begins once every individual in the population has identified their gender orientation, as outlined above. I now proceed to describe the matching procedures, starting with some formal definitions and assumptions.

\begin{quote}
    {\definition{\textbf{Matching Preference ($\mapsto$)}. let $\mapsto$ be a matching preference of one gender identity toward (left-hand side) another (right-hand side). Matches happen in pairs and only if the two involved parties have matching orientations toward each other.} \label{def:matchorientation}}
\end{quote}
    \begin{quote}
    {\definition{\textbf{Strict Matching Preference ($\mapsto\mapsto$)}}{. let $\mapsto\mapsto$ be a strict matching preference where $o \mapsto\mapsto p$ means $o$'s matching orientation is $p$ and nothing else.}}
\end{quote}

\begin{quote}
    {\assumption{{$0 \mapsto\mapsto 1$ and $1\mapsto\mapsto 0$.}} \label{assumption:binary}}
\end{quote}

\begin{quote}
    {\assumption{{$o\in (0,1)  \mapsto p \in[o-b,o+b]$ ($o$ is nonbinary).}}\label{assumption:nonbinary}}
\end{quote}

Assumption \ref{assumption:binary} establishes that strict matching preferences exist between binary gender identities, such that individuals identifying as 0 exclusively prefer to match with individuals identifying as 1, and vice versa. For Assumption \ref{assumption:nonbinary}, with $b\ge0$ is the nonbinary preference bin size, says that nonbinary identities have a more flexible matching preference with matching preference for gender identities within the neighborhood of their identity with a preference bin size $b$.

Once all individuals translated their gender identity, matching commences accordingly to Definition \ref{def:matchorientation}.\footnote{It is noteworthy to emphasize that this concept of matching orientation is narrowly  defined and need not be restricted to certain sexual and cultural interpretations.} An individual in this environment gains utility from its matching likelihood, and hence attributes to the fitness of their genetic trait (i.e., chromosome). Precisely, let $\xi^z$ be a gender identity, and let $\xi^o$ and $\xi^p$ are two gender identities $o$ and $p$ having matching preferences toward each other (i.e., $o \mapsto p$ and $p \mapsto o$), the fitness $\text{Fit}_i$ of a genetic trait $\xi_i$ of an individual $i$ is computed by the following operation:
\begin{equation}
    \text{Fit}_i=\operatorname{Pr}\left(\text {matched} \mid \xi_i=\xi^o\right)=\left\{\begin{array}{c}
1 \text { if } 0<N({\xi^o}) \leq N({\xi^p}) \\
\frac{N({\xi^p})}{N({\xi o})} \text { if } 0<N({\xi^p})<N({\xi^o}) \\
0 \text { otherwise (no match) }
\end{array}\right.\text{, for } o \mapsto p \text{ and } p \mapsto o, \label{eq:fit}
\end{equation}

where $N{(\xi^o)}$ denotes the number of gender identity $o$ in the population. The fitness is maximized at 1 when sufficient matching partners exist, declines proportionally when matches are scarce, and becomes zero if no match is available.

\subsection{The evolutionary algorithm}
\begin{figure}[H]
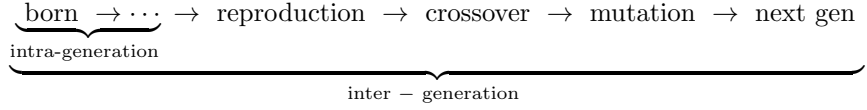

    \centering
\begin{equation*}
    \underbrace{\underbrace{\text { born } \rightarrow \cdots}_{\text {intra-generation }} \rightarrow \text { reproduction } \rightarrow \text { crossover } \rightarrow \text { mutation } \rightarrow \text { next gen }}_{\text {inter }- \text { generation }}
\end{equation*}   
\caption{Inter-generation activities}
    \label{fig:intergen}
\end{figure}

Once matching and fitness computations are completed, the generation concludes, and the current pool of chromosomes undergoes evolutionary procedures to be passed on to the next generation. These procedures—\textit{reproduction}, \textit{crossover}, and \textit{mutation}—yield a new pool of individuals with genetic traits better suited to the environment. The sequence of these processes is depicted in Figure \ref{fig:intergen}.

It is important to note that the concept of a new generation here refers specifically to learning outcomes within the evolutionary algorithm framework, rather than biological reproduction involving mating, emotions, or childbirth. This clarification ensures the focus remains on the algorithm’s mechanics without adding dimensions. In this model, the $type$-chromosome string of an individual remains entirely stochastic (e.g., endowed by nature), and evolutionary processes are applied exclusively to the $\alpha$-chromosome and $\beta$-chromosome. The $type$-chromosome for each new individual is randomly generated at the start of every generation.

\paragraph{Reproduction} This procedure probabilistically selects chromosomes with higher fitness for the next generation. Specifically, $N$ binary tournaments are conducted, where in each iteration, two chromosome strings are randomly selected with replacement from the pool to compete. The chromosome with the higher fitness, as computed by Equation \eqref{eq:fit}, wins the tournament and is copied into the candidate pool for the next generation. This process generates a set of $N$ binary strings with an average fitness that is higher than the previous generation.

\paragraph{Crossover} This operation allows the chromosomes resulting from reproduction to probabilistically meet and exchange genetic traits. Specifically, two binary strings are randomly selected without replacement to ``meet." A random integer 
 $r \in [1,l-1]$; then $\alpha$-chromosome and $\beta$-chromosome parts of the two chromosomes are cut at the $r^{th}$ bit, and then the right-hand side bits of the cut are swapped among two chromosomes with some small positive probability $p^{cross}$. Here is an illustrating example, for chromosomes $j$ and $k$, and $r=6$

\begin{equation*}
    \begin{aligned}
& \underbrace{010010 \mid 0001}_{\alpha_j} \underbrace{110110 \mid 111}_{{\beta_j}}  \\
& \underbrace{011110 \mid 1110}_{\alpha_k} \underbrace{000110 \mid 0010}_{\beta_k} 
\end{aligned}
\end{equation*}

With probability $1-p^{cross}$, there is no change to these two strings. With probability $p^{cross}$, the two strings exchange the bit sections.

\begin{equation*}
    \begin{aligned}
& \underbrace{0100101110}_\alpha \underbrace{1101100010}_\beta \\
& \underbrace{01111000001}_\alpha \underbrace{0001101111}_\beta
\end{aligned}
\end{equation*}

The procedure repeats for another $\frac{N}{2}-1$ times, and returns $N$ new offspring strings.

\paragraph{Mutation} or experimentation is the final genetic operator. All $N$ strings created from reproduction and crossover are subject to the following mutation operation. Each bit $b=0,1$ in a string becomes $1-b$ with some small positive probability $p^{mut}$, with probability $1-p^{mut}$ the bit remains the same.

\section{Game-theoretical predictions: equilibria} \label{sec:gametheory}
\begin{table}[ht]
    \centering
\large\begin{tabular}{|c|c|c|}
\hline & binary & nonbinary \\
\hline binary & $(0.5,0.5)$ & $(0,0)$ \\
\hline nonbinary & $(0,0)$ & $(\Phi, \Phi)$ \\
\hline
\end{tabular}
    \caption{The $2\times 2$ game}
    \label{tab:22game}
\end{table}

In this section, I analytically explore the potential equilibria of the above environment to obtain a benchmark for my computational analysis. For this exercise, I simplify the environment to a population of two individuals, I1 and I2 being row-player and column-player, respectively. Both I1 and I2’s gender set can be either binary or nonbinary. For simplicity, I let I1 and I2 have complete maneurver over its gender identity. The environment boils down to a $2\times 2$ game between the I1 and I2. Table \ref{tab:22game} shows the normal-form game matrix.  The two individuals obtain payoffs from the probability of matching given their chosen gender set. Therefore, when I1 and I2 identify as different gender sets, the payoffs are $(0,0)$. When they both identify as gender identity of binary nature, the expected payoffs are $(0.5,0.5)$. For nonbinary gender identity, the expected payoffs are $(\Phi,\Phi)$, where $\Phi \in [0,1]$ and $\Phi$ increases in the size of the nonbinary preference bin $b$.\footnote{When the preference bin size $b=0$, i.e., a specific nonbinary identity $\in (0,1)$ has matching preference toward itself, the probability of matching is zero, i.e., $\Phi=0$. The reason being the nonbinary identity is drawn from the uniform distribution $[0,1]$, with the probability of drawing two identical nonbinary gender identity being virtually zero.}

{Coincidentally and yet amusingly, this game between I1 and I2 shares similar attributes with the two-player coordination game \textit{battle of the sexes}.}
It is straightforward to see that there are two pure Nash equilibria $\textit{(binary,binary)}$ and $\textit{(nonbinary,nonbinary)}$ in the stage game for all $\Phi \in [0,1]$. These equilibria suggests two stable coordination points, where both players consistently coordinate on either the binary or nonbinary identity.

I now consider the potential dynamics of the outcomes of the game when it is infinitely repeated. First, I consider a refinement of the Nash equilibrium concept in the repeated game: subgame perfect equilibrium (SPE). SPE ensures that strategies form a Nash equilibrium in every subgame, making it essential for analyzing consistent and credible outcomes in the repeated version of the stage game.

\begin{quote}
    {\proposition{ for all possible values of $\Phi$ (i.e., $\Phi \in [0,1]$), in the infinitely repeated game described in Table \ref{tab:22game}, there is a continuum of subgame perfect equilibria with each corresponding to a unique sequence of $\textit{(binary,binary)}$ and $\textit{(nonbinary,nonbinary)}$.  Proof: see Appendix 6.2. $\square$} \label{prop:speprop}}
\end{quote}

Proposition \ref{prop:speprop} implies that in the repeated game, a wide range of equilibrium outcomes can emerge, with the population alternating between binary and nonbinary states, reflecting and ensuring the existence of both equilibria when time is involved. However, this does not say anything about the transitions between the two equilibria. To shed light on the potential transition dynamics between the two equilibria in the stage game, in any subgame perfect equilibrium, I consider a refinement of the Nash equilibrium concept; trembling-hand perfect equilibrium.

\begin{quote}
    {\definition{\textbf{THPE}. a Nash equilibrium is a trembling-hand perfect equilibrium (THPE) if it is robust against small mistakes made by any players. Precisely, a THPE has each of all strategies remains to be the best response even with some arbitrarily small mistake in the opponents’ actions (formal definition in appendix 6.2).}}
\end{quote}

I consider THPE due to its similarity to the mutation procedure. One can loosely regard mutation as a “mistake” in modifying the biological interpreter (e.g., $\alpha$- and $\beta$-chromosome).

 \begin{quote}
     {\proposition{$\textit{(binary,binary)}$ is a trembling-hand perfect equilibrium for all $\Phi \ge 0$. \textit{(nonbinary,nonbinary)} are trembling-hand perfect only if $\Phi>0$. Proof: see appendix 6.2. $\square$} \label{prop:thpeprop}}
 \end{quote}

Proposition \ref{prop:thpeprop} implies that in the limit where the size of the nonbinary bin approaches zero, only $\textit{(binary,binary)}$ is trembling-hand perfect. When $\Phi>0$ (i.e., the nonbinary preference bin size $b$ is sufficiently large), \textit{(nonbinary,nonbinary)} is THPE and becomes robust against small mutation occurrences. In other words, the binary-only equilibrium is universally stable, as it remains robust under all conditions. In contrast, the nonbinary-only equilibrium requires a minimum level of diversity to achieve stability, making it less robust in comparison. This suggests that binary identities are more resilient in environments with low diversity, while nonbinary identities depend on sufficient diversity to persist as a stable outcome. For the following analysis, we let $b$ be arbitrarily small but non-zero.

Overall, the game-theoretical standpoint suggests that both binary-only and nonbinary-only world are both Nash equilibria in the stage game and are locally stable in the infinitely repeated game. Taken together, in an infinitely repeated game, if both individuals are at $\textit{(binary,binary)}$, then there is no transition to $\textit{(nonbinary,nonbinary)}$, and vice versa. Keep this in mind, I now turn to the computerized experiments' results from the GA.

\section{Genetic algorithm simulations} \label{sec:findings}
\begin{figure}[ht]
    \centering
    \includegraphics[scale=0.5]{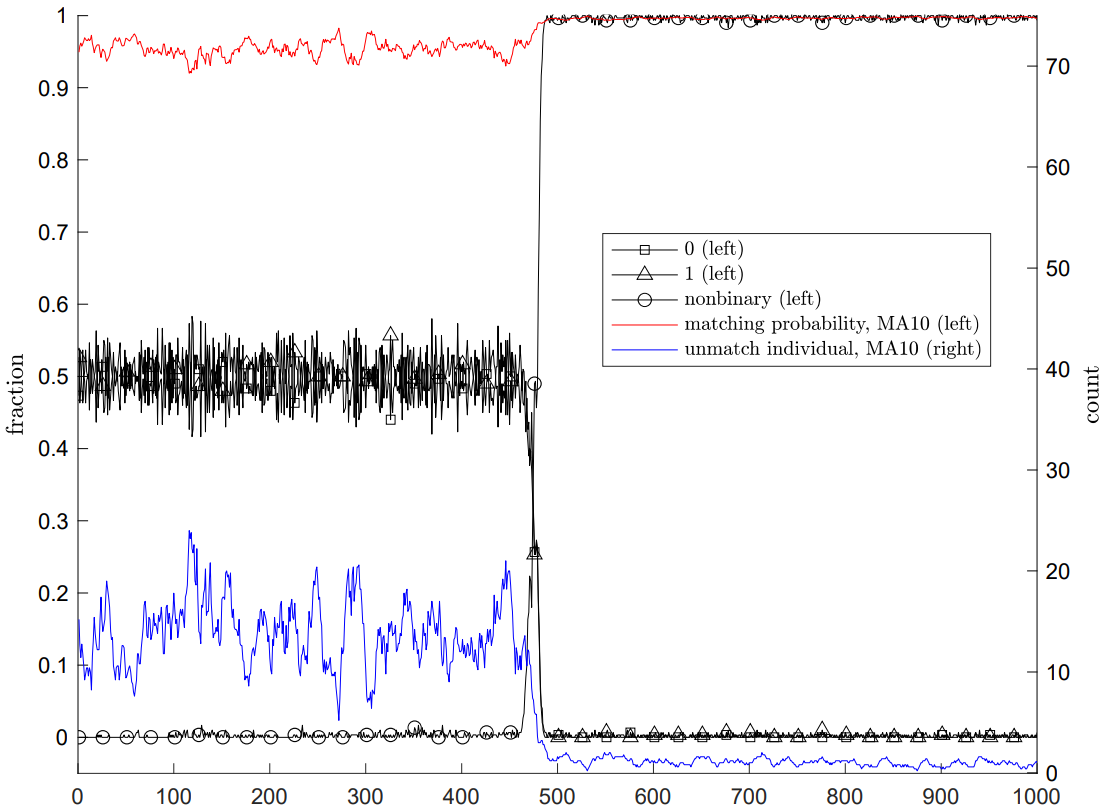}
    \caption{Divergence and convergence to nonbinary gender identity in binary origin world}
    \label{fig:identityconvergence}
\end{figure}
\paragraph{Calibration and simulation protocol} There are four GA parameters. I let the chromosome part length $l=10$, therefore a chromosome has length $3\times l=30$. For the mutation and crossover probabilities, I consider small values, which are consistently in line with the THPE definition. Specifically, $p^{mut}=p^{cross}=0.001$. For the generation size, I let $N=300$, which is standard in the adaptive learning literature of economic decision-making (see, for instance, \cite{arifovic2000evolutionary}).\footnote{I conduct as well a grid of the GA parameters with ranging $l$,$p^{mut}$,$p^{cross}$,$N$; which produce virtually identical output to what I present in the main text, and therefore the reported result is robust to parameters' sensitivity. The results with different GA parameters are available upon request.} As the benchmark analysis focuses on the potential transition from binary to nonbinary identity, I assume that the initial population of individuals is endowed with $\alpha$- and $\beta$-chromosome such that the support of gender identity $\xi$ is $\{0,1\}$, i.e., $\xi \sim \text{Bernoulli}(1/2)$. In other words, the first generation of individuals all identify as binary gender.

A simulation lasts 1000 periods and I conduct 1000 simulations, I then collect the obtained data on individuals' gender identity and the resulting matching outcome.

\paragraph{Results} in Figure \ref{fig:identityconvergence}, I report the median series of the 1000 simulations for gender identities, matching probabilities and the number of unmatched individuals.

The evolutionary dynamics of gender identity, as shown in Figure \ref{fig:identityconvergence}, reveal distinct phases of transition and stabilization.
In roughly half of the simulation periods, the proportions of individuals identifying with binary genders remain the dominant majority, while the proportion of nonbinary identities is negligible, i.e., the minority. This reflects the initial dominance of binary identities, and any ``deviation" to nonbinary identity is not rewarding and reproductive.

Around the midpoint of the simulation, a sharp transition occurs as the proportion of individuals identifying as nonbinary rises significantly, while the proportions of binary identities decline symmetrically. This shift corresponds to a noticeable increase in the overall matching probability, indicating that nonbinary identities improve coordination and matching efficiency due to their greater flexibility.

By the end of the simulation, the population converges predominantly to nonbinary identities, with binary identities nearly disappearing. Simultaneously, the number of unmatched individuals stabilizes at a minimal level, confirming that the dominance of nonbinary identities enhances overall matching outcomes and reduces mismatches.

These results suggest that while binary identities can coexist initially, the adaptive advantages of nonbinary identities, such as improved flexibility in matching, ultimately drive their dominance, supporting the hypothesis that nonbinary identity represents an evolutionarily attractive equilibrium.


\section{Concluding remarks}\label{sec:conclusion}
This study explores the evolutionary dynamics of gender identity, providing a theoretical framework that connects game theory, adaptive learning, and social identity. By introducing a genetic learning algorithm, this paper demonstrates how the transition from binary to nonbinary gender identity can occur over time, driven by the superior flexibility and adaptability of nonbinary identities in enhancing matching outcomes. The findings contribute to our understanding of how diverse social structures emerge and stabilize within varying economic and cultural environments.

The results align with broader patterns observed in identity economics, where adaptive behaviors often drive transitions toward more inclusive and diverse states. This work also highlights the importance of dynamic processes in equilibrium selection, emphasizing that learning and experimentation play pivotal roles in shaping identity distributions. Similar to broader economic development processes, the evolutionary pathway outlined here suggests that nonbinary identities are not only an equilibrium outcome but also an attractor in environments that favor greater social coordination and flexibility.

Importantly, this framework captures critical elements of identity evolution without overcomplicating the model with factors such as cultural norms or emotional drivers, making it a parsimonious yet robust tool for studying identity dynamics. Future research could extend this work by examining the influence of external shocks, institutional factors, or cross-population interactions on identity evolution. These extensions could provide further insights into how social policies or cultural shifts impact the trajectory of identity systems.

Ultimately, this paper underscores the dynamic and adaptive nature of social identities, offering both theoretical insights and practical tools to analyze the interplay between individual behavior and societal outcomes. By bridging theoretical and computational approaches, this research enhances our understanding of identity as a foundational component of economic and social systems.

\begin{appendix}
    \section*{Appendix}
\section{Properties of the biological interpreter}
Let a random variable $\xi \sim \text{Beta}(1/\lambda(\alpha) ,1/\lambda(\beta) )$, I state without proof that if $1/\lambda(\alpha) =1/\lambda(\beta)  \rightarrow 0$, then $\xi$ approaches in the limit to the \text{Bernoulli} distribution that spikes at $0$ and $1$ with probability 1/2 each, and zero probability elsewhere. On the other hand, if $1/\lambda(\alpha) =1/\lambda(\beta) =1$ then $\xi$ has a uniform $[0,1]$ distribution.

\section{Equilibria in the simplified 2×2 game}

\begin{quote}
    {\example{\textbf{Proposition \ref{prop:speprop}}. for all values of $\Phi \in [0,1]$, both $\textit{(binary,binary)}$ and $\textit{(nonbinary,nonbinary)}$ are stage game Nash equilibria. In the infinitely repeated game, for any strategy profile made of a unique sequence of $\textit{(binary,binary)}$ and $\textit{(nonbinary,nonbinary)}$, there is no beneficial unilateral deviation in any period for any players. Then any sequence of $\textit{(binary,binary)}$ and $\textit{(nonbinary,nonbinary)}$ is a subgame perfect equilibrium. This result is derived from the one-stage deviation principle proposition (\cite{fudenberg1991game}). $\square$}}
\end{quote}

    \begin{quote}
        {\definition{\textbf{Formal THPE}. a Nash equilibrium in a normal form game $\Gamma_N$,  $\sigma^*$, is a trembling-hand perfect equilibrium (THPE) if there exists a sequence of fully mixed strategy $\sigma^k$ such that $\sigma^k\rightarrow \sigma^*$ as $\rightarrow \infty$ and for all $i$ and $k$, $\sigma_i^*$ is a best response to $\sigma_{-i}^k$ (\cite{fudenberg1991game}).}\label{def:formalthpe}}
    \end{quote}

\begin{quote}
    {\example{\textbf{Proposition \ref{prop:thpeprop}}. }call the Nash equilibrium $\textit{(binary,binary)}$, $$\sigma^*=\textit{(binary,binary)}=(\sigma_{I1}^*,\sigma_{I2}^* ).$$
For any positive integer $k$, let I1’s strategy be the following mixed form:
$$\sigma_{I1}^k=(binary,(1-1/(k+3));nonbinary,1/(k+3)),$$

and I2’s strategy be:
$$\sigma_{I2}^k=(binary,(1-1/(k+3));nonbinary,1/(k+3)).$$
Then,
$$(\sigma_{I1}^k,\sigma_{I2}^k )\rightarrow (\sigma_{I1}^*,\sigma_{I2}^* )=\textit{(binary,binary)} \text{ as }  k\rightarrow \infty.$$

I1’s expected payoff when playing \textit{binary}, a Nash equilibrium strategy (i.e., $\sigma_{I1}^*$), against I2’s $\sigma_{I2}^k$ is:
$$E(\text{pay}_{I1} |(\sigma_{I1}^*,\sigma_{I2}^k )=1/2(1-1/(k+3)).$$

I1’s expected payoff when playing nonbinary is:
$$E(\text{pay}_{I1} |(\sigma_{I1}^{-*},\sigma_{I2}^k )=1/(k+3).$$

It is straightforward to show that:

$$E\left(\text { pay }_{I 1} \mid\left(\sigma_{I 1}^*, \sigma_{I 2}^k\right)\right)>E\left(\text { pay }_{I 1} \mid\left(\sigma_{I 1}^{-*}, \sigma_{I 2}^k\right)\right) \text { for all } k>0.$$

Thus $\sigma_{I1}^*$ is the best response to $\sigma_{I2}^k$ for all $k$, and by symmetry, $\sigma_{I2}^*$ is also the best response to $\sigma_{I1}^k$ for all $k$. Then $\textit{(binary,binary)}$ is a THPE per definition \ref{def:formalthpe}.

Analogously, $\textit{(nonbinary,nonbinary)}$ is also a THPE only if $\Phi>0$.

Alternatively, since the simplified game is a 2-player game, one can use claim 2.28 in \cite{tamus2018lecture}, which shows that if a Nash equilibrium $\sigma^*$ in a 2-player game $\Gamma_N$ with $\sigma_i^*$ is not weakly dominated for all $i$, then $\sigma^*$ is a THPE. It is straightforward to see that neither $binary$ nor $nonbinary$ are weakly dominated for both I1 and I2. $\square$
}
\end{quote}

\section{Genetic algorithm program outline}
\begin{algorithm}[H]
\caption{Genetic algorithm simulations}
\label{alg:evolution}
\begin{algorithmic}[1]
\State \textbf{Initialize Parameters:} Set periods, agents, chromosome length, mutation, crossover probabilities, and number of replications.

\For{each replication}
    \State \textbf{Initialize Population.}

    \For{$\text{t} \in [0, \text{periods}]$}
        \State \textbf{Decode and Normalize Individuals' Chromosomes.}

        \State \textbf{Gender Identity Translation.}

        \State \textbf{Compute Fitness from Matching.}

        \State \textbf{Genetic Algorithm Operations.} Reproduction, crossover, mutation.
        \State \textbf{Track and Store Metrics.} Gender identities, matching attributes.
    \EndFor
    \State \textbf{Store Data.}
\EndFor

\end{algorithmic}
\end{algorithm}

\end{appendix}


\end{document}